\documentclass[12pt]{article}
\usepackage{epsfig,amsmath,amssymb,array,dcolumn, subfigure,rotating}

\begin{document}

\def\ul#1#2{\textstyle{\frac#1#2}}


\title{Universal Thermal Radiation Drag on Neutral Objects}

\author{Vanik Mkrtchian \dag, V. Adrian Parsegian \ddag, Rudi Podgornik
\ddag $\bullet$, Wayne M. Saslow \S }


\maketitle

\noindent
\dag Institute for Physical Research, Armenian Academy of Sciences,
378410 Ashtarak-2, Armenia \\
\ddag Laboratory of Structural and Physical Biology, National
Institutes of Health, Bethesda, MD \\
$\bullet$ Faculty of  Mathematics
and Physics, University of
Ljubljana, Jadranska 19, SI-1000 Ljubljana,
Slovenia \\
\S Department of Physics, Texas A\&M University, College
Station, Texas 77840-4242

\begin{abstract}
\small
We compute the force on a small neutral polarizable object
moving at velocity $\vec v$ relative to a photon gas equilibrated at
a temperature $T$ We find a drag force linear in $\vec v$.  Its
physical basis is identical to
that in recent formulations of the dissipative component of the
Casimir force. We estimate the strength of this universal Casimir
drag force for different dielectric response functions and
comment  on its relevance in various contexts.

\end{abstract}

PACS numbers:
72.20.-i, 72.20 Jv, 72.40.+w, 73.50.Pz


The residual drag force on an AFM tip close to, but not in direct
contact with a substrate {\sl in vacuo} raises an important and
fundamental question on the origin of non-contact friction
\cite{Giess}.   Other experimental techniques also
are sensitive to non-contact friction \cite{Gramila}. Since it became  clear that the Casimir effect, being {\sl par excellence} a non-contact phenomenon, can lead to a dissipative drag (see \cite{Pen} and references therein), it has become a primary focus of theoretical research \cite{Kardar}.  Such  Casimir  dissipative drag  occurs when electromagnetic field fluctuations equilibrate in a specific reference frame, relative to which another system (e.g. a dielectric or  a conducting  body) is in uniform motion \cite{Pen, Vanik, Ramin}. The difference between the frames of reference in the dissipative Casimir effect can be due to relative motion of {\sl different} bodies, as for two conducting plates with relative motion in the  parallel direction, or for a neutral body moving relative to a conducting plate  \cite{AE1, Liebsch, TW, VPpart, Pen, VP1}.  In these cases the radiation equilibrates in one of the plates, and the friction depends upon the proximity of the other one.

We show here that such
friction can also come about when a {\sl single} body moves relative
to a thermal bath of the electromagnetic field excitations, such as those
between the walls of
an oven or in the cosmic microwave  background.  The friction
has no position dependence, {\sl i.e.} it is spatially homogeneous.
The consequence is a
universal dissipative drag acting on all matter in relative motion
with respect to a thermalized photon gas. To estimate  the magnitude of this universal drag we evaluate it as a function of the dominant
frequency of the electromagnetic response of the body for dielectrics
and conductors.

Consider the Lorentz
force 
\begin{equation}
	\vec F = \int d^3\vec r \left( \rho \vec E + \vec j
\times \vec B
\right).
\label{eq.1}
\end{equation}
on a dielectric in a field $\vec E(\vec r,
t), ~\vec B(\vec r, t)$. The charge and the
current densities in the dielectric are set by the
polarization, $\vec P(\vec r, t)$, such that $\rho (\vec r, t) = -
\vec\nabla \cdot \vec P$
and $\vec j(\vec r, t) = \partial_{t} \vec
P$, where $\vec P, ~ \vec j$ obviously obey the
continuity equation $\partial_t\rho+\vec\nabla\cdot\vec j=0$.  Assume that the response of the matter to the field is both  linear as well as spatially local.  Thus we write
\begin{equation}
	\vec P (\vec r, t) = \epsilon_{0} \int \chi_e (t -
t') \vec E (\vec
r, t') dt',
\label{eq.1.1}
\end{equation}
in MKS units,
where the dielectric susceptibility $\chi_e$ is dimensionless. In the
frequency domain $\chi_e (\omega)$ is equal to
$(\epsilon (\omega) - 1)/\epsilon_0$ for condensed matter.  For weakly interacting molecules or
atoms $\chi_e= \rho_N \alpha_m$, with $\rho_N$ the atomic or molecular number density of the medium and $\alpha_m$ the polarizability  of the single molecule or atom.

With the polarization proportional to the field, the force is
bilinear in the field.  This bilinearity holds for objects moving at
arbitrary non-relativistic velocities relative to the frame of
reference of the thermal bath.   We require the thermal average of the force acting on a moving body in unbounded space filled with radiation at rest, working in the reference frame where the
particle is instantaneously at rest and the photon gas moves at velocity $\vec v$. The average force is  obtained in terms of the thermal averages of the Fourier components of the
field correlations.  In unbounded space the Fourier components of the
polarization and electric fields
are
\begin{equation}
\vec P(\vec k,
\omega), \vec E(\vec k, \omega)
=\int dt \int d^3\vec r ~\vec P(\vec r, t),
\vec E(\vec r, t)~
e^{-i(\vec k\cdot\vec r-i\omega t)}
\label{eq.2}
\end{equation}
leading to the following form of the thermal
average $\left<
..\right>$ of the Lorentz force
\begin{eqnarray}
\kern-20pt & &\left< \vec F \right> = - i \epsilon_{0} \int d^3\vec
r \int \frac{d^{3} \vec k d \omega}{(2\pi)^{4}} \int \frac{d^{3} \vec k'
d \omega'}{(2\pi)^{4}} e^{-i(\vec k + \vec k') \cdot\vec r + i
(\omega + \omega') t } ~\chi_e(\omega) \times \nonumber\\
	& & \times \left<
\frac{\omega}{\omega'} \left( \vec k' \left( \vec
E (\vec k, \omega) \vec
E (\vec k', \omega') \right) - \left( \vec k'
\cdot \vec E (\vec k,
\omega)\right) \vec E (\vec k', \omega')\right)
\right>,
\nonumber\\
\label{eq.2.1}
\end{eqnarray}
where we took note of the fact that
in empty space the electric field
has no sources.  A slight generalization of the
standard expression (\cite{LL-SM2} Eq.
77.12) for the thermal average of
the correlator of the electric field vectors yields
\begin{equation}
\left<
E_i(\vec k, \omega) E_j(\vec k', \omega') \right> =
(2\pi)^4\delta(\vec
k+\vec k')\delta(\omega+\omega')\left< E_i E_j
\right>_{\vec k, \omega}.
\label{eq.3}
\end{equation}
with
\begin{equation}
<E_i E_j>_{\vec k,
\omega}={2\pi^2\hbar\over {\epsilon_{0} k}}
\left(\frac{\omega^2}{c^2}\delta_{ij}-k_ik_j\right)
\left[ \delta({\omega\over
c}-k)-\delta({\omega\over c}+k)\right]
\left( 1+2n(\omega, \vec k)\right),
\label{eq.4}
\end{equation}
where $k=|\vec k|$ and
\begin{equation}
n(\omega, \vec k)=\frac{1}{e^{\beta\hbar(\omega-\vec
k\cdot\vec v)}-1}
\nonumber
\end{equation}
is the Bose occupation number
for a photon distribution at
temperature $T$, moving with velocity $\vec
v$, and $\beta=(k_B
T)^{-1}$. Eq.~\ref{eq.4} is obtained by generalizing
the  fluctuation-dissipation theorem to a translationally invariant
system, and taking a Gibbs distribution corresponding to a photon gas
moving with velocity $\vec v$, just as done for excitations in superfluidity \cite{LL-SM1}.
\begin{figure}[h]
\begin{center}
\epsfxsize=16cm
    \epsfig{file=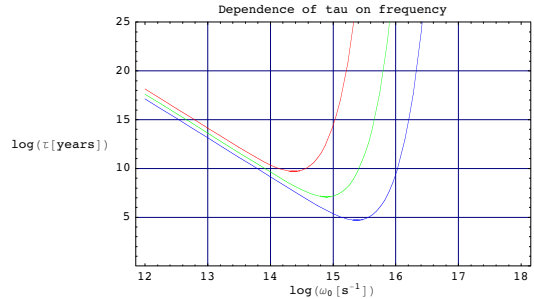, width=8cm}
\end{center}
\caption{Dependence of the logarithm of the
relaxation time in years on the
logarithm of the primary relaxation frequency ($\omega_0$ in Eq.
\ref{eq.9}) for three
different temperatures, $T=300$~K, $T=1000$~K, $T=3000$~K
(upper, middle and lower curve). The frequency of the minimum is
given by $\hbar \omega_0 =  ~5.9694 ~ k_B T$.}
\label{fig1}
\end{figure}
From Eq. \ref{eq.4} we end up with the following average of the Lorentz force
\begin{equation}
	\left< \vec F \right> = - i
\epsilon_{0} \int d^3\vec r \int
\frac{d^{3} \vec k ~d \omega}{(2\pi)^{4}}
~\chi_e(\omega)~ \vec k
\left< \vec E \cdot \vec E  \right>_{\vec k,
\omega}.
\label{eq.5}
\end{equation}
Substitution of (4) in
(5) with $\vec v=  0$ gives an integral
over $\vec k$ that, by symmetry, is zero.  To obtain a non-zero
result it is convenient to first
eliminate the delta functions by
integrating over $\omega$ and then
expanding in powers of $\vec v$
(as done to obtain the normal fluid
density in the theory of
superfluidity). Integration over $\omega$ then
yields
\begin{eqnarray}
<\vec F> &=& -i~ 2\pi\hbar c ~V~\int{d^3\vec k\over
(2\pi)^3} ~k \vec
k~
\chi_e(ck)\left( 1+2n(ck, \vec k)\right) +
\nonumber\\
& & i~ 2\pi\hbar c ~V~\int{d^3\vec k\over (2\pi)^3} ~k \vec k~
\chi_e(-ck)\left(1+2n(-ck, \vec
k)\right),
\label{force}
\end{eqnarray}
where we have taken the dielectric
to be homogeneous so the integral
over its volume simply gives $V$.

The same equation also can be obtained from energy loss considerations,
following the arguments of Volokitin and Persson \cite{VPpart}. One starts
from the  dissipation rate of the energy in the rest frame of the
fluctuating field, 
\begin{equation}
\frac{dW}{dt} =  \int
d^3\vec r \left( \vec j + \rho \vec v  \right)
\cdot \vec E =
\frac{dW_0}{dt}  - \vec F \cdot \vec v.
\end{equation}
Here we take into
account that in the rest frame of the thermalized
photon gas the total
electric current is given {\sl via} the
linearized Lorentz form $\vec j
\longrightarrow \vec j + \rho \vec
v$. Apart from the heat production $\frac{dW_0}{dt}$ in
the frame of the body,
Eq. \ref{force} is reproduced
immediately.
\begin{figure}[h]
\begin{center}
\epsfxsize=16cm
    \epsfig{file=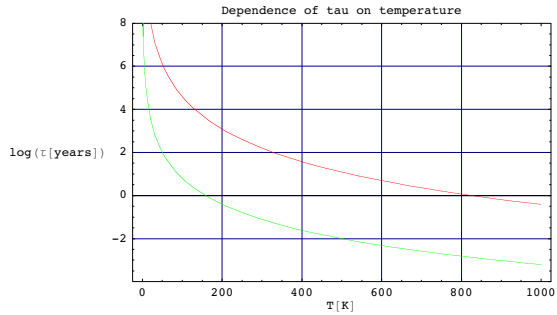, width=8cm}
\end{center}
\caption{Dependence of the relaxation time in
years on temperature. For the dielectric response (upper curve) we plot the minimal
relaxation time obtained for the relaxation frequency that is
proportional to the first Matsubara frequency. We take $\rho_M$ to
be the density of water and the dimensionless measure of atomic polarizability
$\alpha_0 =(N/V)\epsilon\alpha_m \approx 1$. Numerically this
value corresponds to a single particle of molecular polarizability $\epsilon_0 \alpha_m
\approx 1.0 \times 10^{-30}$~m$^3$ and a mass equal to the proton mass $1.67 \times
10^{-27}$~kg. For the metallic response (lower curve) we take the
characteristic time $\tau_0 = \frac{\epsilon_0}{\sigma} \sim 10^{-18}$~s,
well within the range of conductivities of simple metals. }
\label{fig2}
\end{figure}

We now write, in a standard way, $\chi_e(x) =
\chi_e'(x) + i \chi_e''(x)$, noting that the real part is an even and the
imaginary part is an odd function of the argument. The expression for the
force can now be obtained in the following form
\begin{eqnarray}
\kern-30pt <\vec F> = -i~ \frac{4\pi\hbar c
V}{(2\pi)^3}\int{d^3\vec
k} ~k \vec k\!\!\!&&\!\!\!\left[ \chi_e'(ck)
\left(n(ck, \vec k) -
n(-ck, \vec k)\right) +  \right. \nonumber\\
& &
\left. + ~i~\chi_e''(ck) \left( 1 + n(ck, \vec k) + n(-ck, \vec
k)\right)\right].
\end{eqnarray}
Expanding the Bose function to the lowest order in velocity, we have
$ n(ck, \vec k) - n(-ck, \vec k) = \coth{\ul12
\beta \hbar c k} +
{\cal O}(v^2)$ and $ 1 + n(ck, \vec k) + n(-ck, \vec k)
= \ul12 {\rm
csch}^2{\ul12 \beta \hbar c k}~\left(\beta \hbar (\vec k
\cdot \vec
v)\right) + {\cal O}(v^3)$. Placed in Eq. \ref{force}, the $\vec k$-space
integral over the term in $\chi_e'(ck)$ is zero,  by symmetry.  The
remaining term then yields
\begin{equation}
<\vec F> = (8\pi \beta
\hbar^2 c) ~V~\int{d^3\vec k\over (2\pi)^3}
~\frac{k~\vec k (\vec k\cdot\vec v)~ \chi_e''(ck)}{ {\rm sinh}^2{\ul12
\beta \hbar c  k} },
\label{eq.6}
\end{equation}
Performing the angular integral over $d^3 \vec
k$ gives a factor of $4\pi/3$. Reverting again to $\omega$ by
substituting $k = \omega/c$  gives
\begin{equation}
<\vec F> = V~\vec
v~\left({\beta \hbar^{2} \over 3\pi c^5}\right)
~\int_0^\infty d\omega
~\frac{\omega^5 ~ \chi_e''(\omega)}{{\rm
sinh}^{2}(\ul12 \beta \hbar
\omega)}.
\label{eq.7}
\end{equation}
This is the fundamental result of our paper. It states that the EM
field fluctuations exert a drag, proportional in the lowest order to
the velocity, on a particle that moves with respect to the frame of
reference in which the EM field fluctuations are thermalized.  Setting this force to $M\vec v/\tau$, where $M$ is the total mass of  the object and $1/\tau$ is the drag time, and using
$\rho_M=M/V$,  yields the result that
\begin{equation}
{1\over\tau} =
\left({\beta \hbar^{2}\over 3\pi \rho_M
c^5}\right)\int_0^\infty d\omega
~\frac{\omega^5
\chi_e''(\omega)}{\sinh^{2}(\ul12 \beta \hbar \omega)}.
\label{eq.8}
\end{equation}
Regard this fundamental result in
three different  contexts: molecules, dielectric and conducting condensed
matter.  First assume that the dielectric response of the medium
can be characterized by a single sharp absorption line at $\omega_0$.
Because $1\over\tau$ is proportional to $\chi_e''(\omega)$, obviously each of
the absorption lines for a molecule or a dielectric will contribute
additively to the integral in Eq. \ref{eq.8}.
We set $\chi_e''(\omega) =
\alpha_0\delta(\omega/\omega_0-1)$, where
$\alpha_0$ is a constant
describing the strength of the response.
This assumption gives
\begin{equation}
\tau = \left({3\pi n c^5\over \beta
\hbar^{2}}\right)~\frac{\sinh^{2}(\frac{\beta \hbar
\omega_{0}}{2})}{\alpha_{0} \omega_{0}^{6}} = \left({3\pi \rho_M c^5
\hbar^4 \over 2^6 \alpha_{0} (k_B T)^5}\right)~\frac{\sinh^{2}(x)}{
x^6},
\label{eq.9}
\end{equation}
after introducing $x = \ul12 \beta \hbar \omega_0$. In
this form the relaxation time depends strongly on the absorption frequency
$\omega_0$.  It has a minimum at a temperature dependent
frequency (see Fig. \ref{fig1}) that
coincides with the minimum of the function $f(x) =
\frac{\sinh^{2}(x)}{ x^6}$, at $x_m = 2.98$, where  $f(x_m) =
0.137$. Because $f(x)$ is the square of the function
$\frac{\sinh(x)}{ x^3}$, it has a broad minimum with a quartic,
rather than a quadratic, dependence upon the deviation from $x_m$.  Taking
this minimum at $x_m$ into account, the smallest possible relaxation
time can thus be obtained from the above equation in the form
\begin{equation}
\tau = \tau_0 \left(
{T_0 \over T}\right)^5,
\label{findie}
\end{equation}
where $\tau_0 T_0^5
= \left({3\pi f(x_m) \over
2^6}\right)~\frac{\rho_M c^5
\hbar^4}{\alpha_{0} k_B^5}$. At this  minimum the  absorption
frequency $\hbar \omega_0 = ~2 ~x_m ~ k_B T$ is proportional to the first
Matsubara frequency. The temperature dependence of this minimal possible
relaxation time is  given in Fig. \ref{fig2}.

A different formula for
$\tau$ is obtained for metals, which have
constant conductivity at frequencies below the collision time
of their charge-carriers.
In this case $\chi_e(\omega) \approx -\frac{\sigma}{i
\epsilon_0 \omega}$.
Inserting this into Eq. \ref{eq.8}, the inverse
relaxation time for drag
now takes the form
\begin{equation}
{1\over\tau} =
\left({\beta \hbar^{2} \sigma \over 3\pi \rho_M c^5
\epsilon_0} \right)
\int_0^\infty d\omega
~\frac{\omega^4}{\sinh^{2}(\ul12 \beta \hbar
\omega)},
\end{equation}
so that 
\begin{equation}
\tau = \tau_0 \left(  {T_0 \over
T}\right)^4,
\label{finmet}
\end{equation}
with $\tau_0 =
\frac{\epsilon_0}{\sigma}$ and $T_0^4 = \frac{45 c^5
\hbar^3 \rho_M}{16
\pi^2 k_B^4}$. For most common metals the value of
$\tau_0$ is between $10^{-19} - 10^{-17}$~s.  Taking its geometrical average
we obtain the temperature dependence of the relaxation time as
given in Fig. \ref{fig2}.

The times given in the figures are relatively
long,  corresponding to the general weakness of the Casimir interaction and
the even lower strength of the dissipative Casimir effect.
Two circumstances under which such long times might be observable
are ovens and the cosmos.

For ovens, on trapping molecules with
unusually large polarizabilities it might  be possible to observe a
resonance in an ion or atom trap with a quality factor ${\cal Q}$
that is determined by the
dissipative Casimir interaction.  Small metal particles in a high
temperature oven should also be susceptible to the effects of
the Casimir drag.

For the cosmos, it is
believed that hydrogen atoms  condensed from protons and electrons at about
3000~K, and that the  coupling of radiation and matter due to Compton
scattering becomes  ineffective below this condensation temperature
\cite{Peebles}.
However, as is clear from Fig.~2, it should not be  difficult for
molecules to remain coupled to the cosmic microwave background
when the temperature was between about 300~K (and perhaps a bit less)
and the 3000~K condensation temperature.  This
coupling could have an influence on the structure and anisotropies
observed in recent experiments on the cosmic microwave background.
It could also influence the behavior of molecules formed from
the residue of novas and supernovas, and then subject to drag from a
still-hot cosmic microwave (i.e., electromagnetic) background.

The universal thermal drag calculated in this contribution is yet
another important facet of the Casimir effect, which in this context
appears to play a role not only in the static interactions between
dielectrically inhomogeneous bodies but also in the dissipative
dynamics of particles in homogeneous space.

\end{document}